\documentclass{elsart}

\usepackage{amsmath}
\usepackage{amssymb}
\usepackage{graphicx}
\usepackage{dcolumn}
\usepackage{bm}

\def\be{\begin{eqnarray}}
\def\ee{\end{eqnarray}}
\def\ba{\begin{array}}
\def\ea{\end{array}}

\begin{document}

\begin{frontmatter}

\title{Where are the edge-states near the quantum point contacts?
\\ A self-consistent approach.}

\author[l1]{A. Siddiki}
\author[l2]{, E. Cicek}
\ead{engincicek@trakya.edu.tr}
\author[l2]{, D. Eksi}
\author[l2]{, A. I. Mese}
\author[l2]{, S. Aktas}
\author[l3]{ and T. Hakio\u{g}lu}

\address[l1]{Physics Department, Arnold Sommerfeld Center for Theoretical Physics, and Center for NanoScience
Ludwig-Maximilians-Universit\"at M\"unchen, D-80333 Munich,
Germany}
\address[l2]{Trakya University, Department of Physics, 22030 Edirne, Turkey}
\address[l3]{Department of Physics and UNAM Material
Science and Nanotecnology Research Institute Bilkent University, Ankara, 06800
Turkey}

\begin{abstract}
In this work, we calculate the current distribution, in the close
vicinity of the quantum point contacts (QPCs), taking into account
the Coulomb interaction. In the first step, we calculate the bare
confinement potential of a generic QPC and, in the presence of a
perpendicular magnetic field, obtain the positions of the
incompressible edge states (IES) taking into account
electron-electron interaction within the Thomas-Fermi theory of
screening. Using a local version of the Ohm's law, together with a
relevant conductivity model, we also calculate the current
distribution. We observe that, the imposed external current is
confined locally into the incompressible strips. Our calculations
demonstrate that, the inclusion of the electron-electron
interaction, strongly changes the general picture of the transport
through the QPCs.
\end{abstract}
\end{frontmatter}
%

\section{Introduction}
At low temperatures, low-dimensional electron systems manifest
peculiar quantum-transport properties. One of the key elements of
such transport systems are the quantum point contacts (QPC)
constructed on a two-dimensional electron system (2DES). The wide
variety of the experiments concerning QPCs ~\cite{Whraham88:209},
including quantum Hall effect (QHE) based Mach-Zehnder
interferometer (MZI)~\cite{Heiblum03:415,Neder06:016804}, have
attracted many theoreticians to investigate their
electrostatic~\cite{SiddikiMarquardt} and transport
properties~\cite{Meir02:196802,Igor07:qpc2}. However, a realistic
modelling of QPCs that also takes into account the involved
interaction effects is still under debate. The magneto-transport
properties of such narrow constrictions is typically based on the
standard 1DES~\cite{Buettiker88:317}, which relates the
conductance through the structure to its scattering
characteristics, considering typically a hard-wall confinement
potential. The reliability of such non-interacting approaches is
limited, since interactions are inevitable in many cases and plays
a major role in determining the electronic and transport
properties. In order to account for the interactions simplified
models are used with some phenomenological parameters, which is
not always evident whether such a description is sufficient to
reproduce the essential physics.

The QHE based MZI~\cite{Heiblum03:415} has become a central
interest to the community, since it provides the possibility to
infer interaction mechanisms and
dephasing~\cite{Florian04:056805,Samuelson04:02605} between the ES
by achieving extreme contrast interference oscillations. In these
experiments
ESs~\cite{Buettiker88:317,Chklovskii92:4026,siddiki2004,Igor04:3136}
are assumed to behave like optical beams, whereas QPCs simulate
the semi-transparent mirror in its optical counterpart. The
unexpected behavior of interfering electrons, such as
path-length-independent visibility oscillations, is believed to be
related to long range $e-e$ interactions. Thus, the experimental
findings present a clear demonstration of the breakdown of,
commonly used, Landauer's conductance picture away from the linear
regime. Here, we calculate the effective potential in a
self-consistent manner and , in addition, using a local version of
the Ohm's law within and out-of-the-linear-response regime, we
obtain the current distribution near the QPC's. We essentially
show that, in the presence of an IES, the imposed current is
confined to this region, otherwise is distributed classically.
\section{Model, Results and Discussion}
Our aim is to calculate the distribution of the ES within a
interacting model. We start with the bare confinement potential
obtained from the lithographically defined construction, following
Ref's~\cite{SiddikiMarquardt,Davies94:4800}. For a given pattern
of (metallic) gates residing on the surface and the potential
values $V_{\rm g}(x,y,0)$, one can obtain the potential
experienced $V_{\rm ext}(x,y)$ by the 2DES beneath using
semi-analytical scheme~\cite{Davies94:4800} yielding \be
\label{eq:davies}V_{\rm{ext}}({\bf{r}},z)=\frac{1}{\kappa}
\int{\frac{|z|}{2\pi(z^{2}+|{\bf{r}}-{\bf{r'}}|^{2})^{3/2}}}V_{\rm
g}({\bf{r'}},0)d{\bf{r'}}, \ee where $\kappa$ is the dielectric
constant of the hetero-structure ($\approx12.4$ for GaAs/AlGaAs)
and ${\bf{r}}=(x,y)$. Given the external potential in the position
space, it is straight forward to calculate the screened potential
in the momentum space $(q)$ by $ V_{\rm scr}(q)=V_{\rm
ext}(q)/\epsilon(q) \label{eq:vscr}$ using the Thomas-Fermi
dielectric function, $\epsilon(q)=1+1/(a_0|q|)$, where
$a_0/2=a^{\star}_B=\bar{\kappa} \hbar^2 /(m e^2)$ (for GaAs
$a_{\rm B}^{\star} =9.8\,$nm). We use this potential to initialize
the self-consistent scheme described below, to obtain density
$n_{\rm el}(\bf{r})$ and potential distribution in the presence of
a perpendicular magnetic field, $B$, at a finite temperature $T$,
$\mu^{\star}({\bf{r}})$ is SC'ly found. In the absence of an fixed
external current $I$, $\mu^{\star}({\bf{r}})$ is position
independent and is constant all over the sample, which is in turn
determined by the average electron (surface) number density,
$\bar{n_{\rm el}}({\bf{r}})$. In our calculations we set
$\overline{n}_{\rm el}=3.0\cdot10^{11}$ cm$^{-2}$, corresponding
to a Fermi energy $E_{\rm F}\sim 10.7$ meV. Starting from $V_{\rm
scr}({\bf{r}})$ one can obtain the electron density distribution,
within the Thomas-Fermi approximation~\cite{Oh97:13519} (TFA),
from
\be \label{thomas-fermi}
 n_{\rm el}({\bf{r}})=\int dE\,D(E)f\big( [E+V({\bf{r}})-\mu^{\star}]/k_{B}T \big),
\ee
where $D(E)$ is the Gaussian broadened (single-particle) density
of states (DOS) and $f(\xi)=[1+e^{\xi}]^{-1}$ the Fermi function.
The total potential energy of an electron, $V({\bf{r}})=V_{\rm
ext}({\bf{r}})+V_H({\bf{r}})$, differs from the Hartree potential
energy $V_H({\bf{r}})$ by the contribution due to external
potentials and is calculated from \be \label{hartree}
V({\bf{r}})=V_{\rm ext}({\bf{r}}) + \frac{2e^2}{\bar{\kappa}}
\int_{A} \!d{\bf{r'}} K({\bf{r}},{\bf{r'}})\,n_{\rm
el}({\bf{r'}}). \ee For periodic boundary conditions, the kernel
$K({\bf{r}},{\bf{r'}})$ can be found in a well known text book
~\cite{Morse-Feshbach53:1240}, otherwise has to be solved
numerically.

In a classical manner, if a current is driven in $y$ direction in
the presence of a perpendicular $B$ field a Hall potential develops
in the $x$ direction. Therefore the electrochemical potential has to
be modified due to external field in the $y$ direction with $
E(\textbf{r})=\nabla \mu^\star(\textbf{r})/e
=\hat{\rho}(\textbf{r})j(\textbf{r})$, for a given resistivity
tensor and boundary conditions, in the thermal-equilibrium
(locally), which brings a new self-consistent loop to our problem.
We calculate the electric field by solving the equation of
continuity under static conditions, $\nabla.j({\bf{r}})=0$, and
$\nabla \times E({\bf{r}})=0$, for a fixed total current,
self-consistently.

\begin{figure}
{\centering
\includegraphics[width=1.0\linewidth]{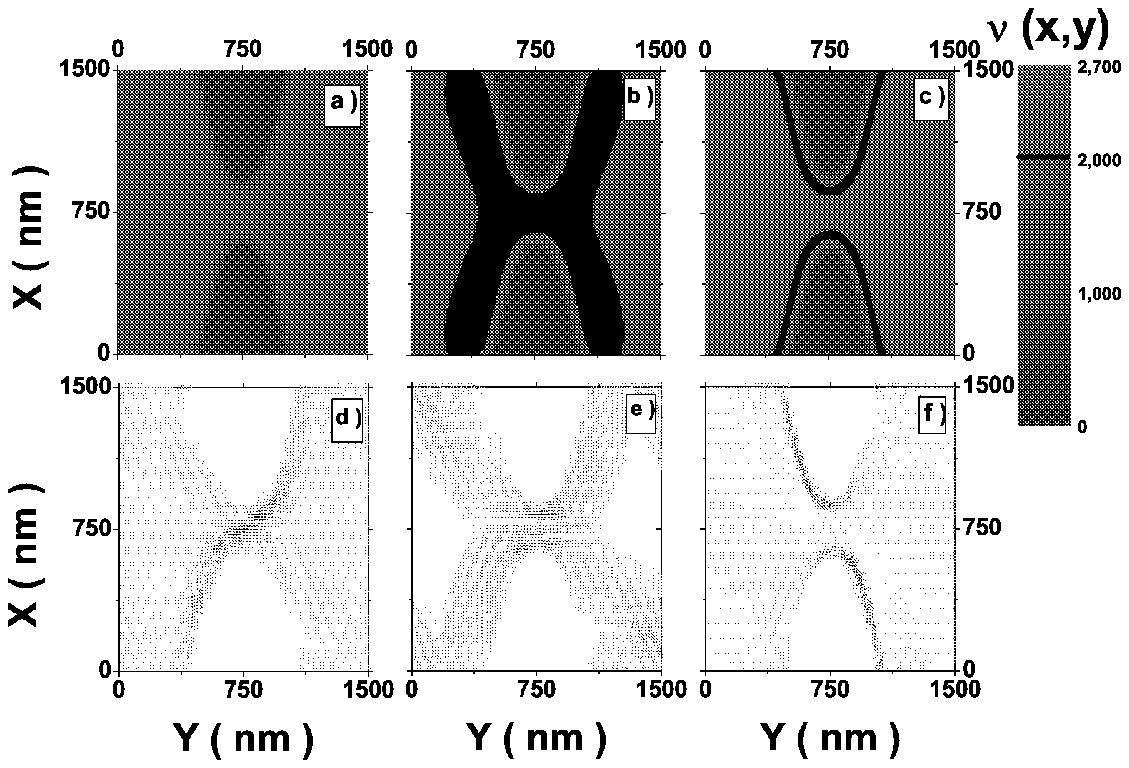}
%
\caption{ \label{fig:fig2} The color coded $\nu(x,y)$ for
$\Omega=1.4,1.24,1.13$ (a-c) and corresponding current distribution
indicated by the arrows. The calculations are done at $\Omega/k_{B}T
\approx 0.025$ for a fixed external current driven in the $y$
direction ($j({\bf r})=j_{0}(0,0.1)$), where the 1D current density
is set to $-0.42\times10^{-2}$ A/m. The gates defining the QPC, are
taken to be $300$ nm apart and biased with $-0.3$ V. The 2DES is
$85$ nm below the surface and unit cell is a square with dimensions
$1500\times1500$ nm$^2$.}}
\end{figure}

As mentioned above, in the single particle model it is believed
that the current is carried by the ballistic 1D
Landauer-B\"uttiker (LB)-ES and the conductance is obtained by the
transmission coefficients. Here we calculate, the electron and
current density self-consistently and observe the different
distributions of the IES under quantum Hall conditions, within the
linear response regime. We will always consider the case, where
$E_F$ is larger than the hight of the barrier at the center of the
QPC, so that the conductance is finite and it is at least equal to
the first Landau energy ($\Omega/2=\hbar \omega_c/2)=eB/2m$). In
Fig.~\ref{fig:fig2}, we selected three representative $B$ field
values, such that (i) the system is almost compressible (a); (ii)
the IES merge at the opening of the QPC (b) and; (iii) the IES
percolate through the constraint (c). From the "classical" current
point of view we observe that, the current biased from bottom is
(almost) homogeneously distributed all over the sample if there
exists no IES inside the constraint, the current passes through
the QPC and ends at the right top contact, mostly (d).
Fig.~\ref{fig:fig2}e, shows us that, the current is confined to
the IES and conductance is quantized, whereas for $\Omega=1.4$ it
is a bit larger than $e^2/h$. These results indicate that, the ESs
present structures inside the QPCs if one models them in a more
realistic scheme rather than as a single point, although the
conductance quantization remains unaffected. Since now we can
calculate the widths of the IES, depending on the magnetic field
and sample structure it is also possible within this model to
obtain the electron velocity inside IES, which may be combined
with a recent work by I. Neder~\cite{Neder07:112}. This work is
based on non-Gaussian noise measurements at the Mach-Zehnder
interference experiments. Their main finding is that the
unexpected visibility oscillations observed, can be explained by
the interaction between the "detector" and "interference" edge
channels. The essential parameters of this work are the electron
velocity at the detector edge channel and the coupling
(interaction) strength between the detector and the interference
channels. We believe that, an extension of our present model to a
realistic system may contribute to the understanding of the
mentioned experiments.

We would like to thank R. R. Gerhardts for his fruitful lectures
on screening theory, enabling us to understand the basics. The
authors acknowledge the support of the Marmaris Institute of
Theoretical and Applied Physics (ITAP), TUBITAK grant $105T110$,
TUBAP-739-754-759, SFB631 and DIP.

\end{document}